\documentclass[11pt, reqno]{article}

\usepackage{jheppub}
\usepackage{amssymb}
\usepackage{amsmath}
\usepackage[usenames,dvipsnames]{xcolor}
\usepackage{epsfig}
\usepackage{dcolumn}
\usepackage{tikz}
\usetikzlibrary{shapes.geometric, arrows}
\usepackage{upgreek}
\usepackage{setspace}
\usepackage{enumitem}
\usepackage{array,multirow,bigdelim,arydshln}
\usepackage{appendix}
\usepackage{xparse}
\usepackage[utf8]{inputenc}
\hypersetup{
	colorlinks,
	urlcolor=Maroon,
	linkcolor=Maroon,
	citecolor=Maroon
}

\NewDocumentCommand{\binomial}{omm}
 {%
  \genfrac(){0pt}{}{#2}{#3}%
  \IfValueT{#1}{_{\!#1}}%
 }
\NewDocumentCommand{\eulerian}{omm}
 {%
  \genfrac<>{0pt}{}{#2}{#3}%
  \IfValueT{#1}{_{\!#1}}%
 }

\newcommand{\tikzmark}[2]{\tikz[overlay,remember picture,baseline] \node [anchor=#2] (#1) {};}

\usepackage{latexsym}
\usepackage{tikz}

\title{Scattering Equations:\\
Real Solutions and Particles on a Line}
\author[a]{Freddy Cachazo,}
\author[a,b]{Sebastian Mizera,}
\author[a,b]{and Guojun Zhang}
\affiliation[a]{Perimeter Institute for Theoretical Physics, Waterloo, ON N2L 2Y5, Canada}
\affiliation[b]{Department of Physics $\&$ Astronomy, University of Waterloo, Waterloo, ON N2L 3G1, Canada}
\emailAdd{fcachazo, smizera, gzhang2@pitp.ca}
\arxivnumber{1609.00008}

\abstract{We find $n(n-3)/2$-dimensional regions of the space of kinematic invariants, where all the solutions to the scattering equations (the core of the CHY formulation of amplitudes) for $n$ massless particles are real. On these regions, the scattering equations are equivalent to the problem of finding stationary points of $n-3$ mutually repelling particles on a finite real interval with appropriate boundary conditions. This identification directly implies that for each of the $(n-3)!$ possible orderings of the $n-3$ particles on the interval, there exists one stable stationary point. Furthermore, restricting to four dimensions, we find that the separation of the solutions into $k\in \{2,3,\ldots ,n-2\}$ sectors naturally matches that of permutations of $n-3$ labels into those with $k-2$ descents. This leads to a physical realization of the combinatorial meaning of the Eulerian numbers.}

\begin{document}
\maketitle
\addtocontents{toc}{\protect\setcounter{tocdepth}{1}}
\def \tr {\nonumber\\}
\def \la  {\langle}
\def \ra {\rangle}
\def\hset{\texttt{h}}
\def\gset{\texttt{g}}
\def\sset{\texttt{s}}
\def \be {\begin{equation}}
\def \ee {\end{equation}}
\def \ba {\begin{eqnarray}}
\def \ea {\end{eqnarray}}
\def \k {\kappa}
\def \h {\hbar}
\def \r {\rho}
\def \l {\lambda}
\def \be {\begin{equation}}
\def \en {\end{equation}}
\def \bes {\begin{eqnarray}}
\def \ens {\end{eqnarray}}
\def \red {\color{Maroon}}

\numberwithin{equation}{section}

\section{Introduction}

The space of kinematic invariants for the scattering of $n$ massless particles has very striking connections to the moduli space of $n$-punctured Riemann spheres \cite{Witten:2003nn,Roiban:2004yf,Berkovits:2004hg,Cachazo:2012da,Cachazo:2012kg,Cachazo:2012pz}. One such connection is independent of the spacetime dimension. This is given by the set of scattering equations \cite{Fairlie:1972zz,RobertsPhD,Fairlie:2008dg,Gross:1987kza,Gross:1987ar,Witten:2004cp,Makeenko:2011dm,Cachazo:2012uq,Cachazo:2013iaa,Cachazo:2013gna,Mason:2013sva}
\be
\sum_{\substack{j=1\\ j\neq i}}^n\frac{s_{ij}}{\sigma_i-\sigma_j} =  0 \quad {\rm for} \quad i\in\{ 1,2,\ldots ,n\}.
\ee
These simple-looking equations possess a rich structure that has attracted considerable attention in the recent years \cite{Kalousios:2013eca,Dolan:2014ega,He:2014wua,Huang:2015yka,Cardona:2015eba,Cardona:2015ouc,Dolan:2015iln,Gomez:2016bmv,Bosma:2016ttj}. Here $s_{ij}$ is the matrix of two-particle Mandelstam invariants. In general dimensions, the $n\times n$ real matrix $s_{ij}$ is only required to be symmetric, have zeros on the diagonal, i.e., $s_{ii}=0$, and have the sum of its rows (columns) vanish. The space of such matrices is $n(n-3)/2$-dimensional and will be denoted by ${\cal K}_n$. In dimensions $d<n-1$ there are extra constraints on the minors of the matrix.

Unitarity and locality restrict the location of possible poles of scattering amplitudes to subspaces of ${\cal K}_n$ where either a single $s_{ij}=0$ (collinear singularities) or, for any subset ${\cal I} \subset \{1,2,\ldots,n\}$ with at least three labels (and less than $n-2$),
\be
\sum_{i,j\,\in\, {\cal I}} s_{ij}  = 0 \quad {\rm with} \quad s_{ij}\neq 0.
\ee
These are called multi-particle singularities. The special property of the scattering equations is that they connect the singularity structure of ${\cal K}_n$ to that of the boundaries of the moduli space of $n$-punctured Riemann sphere \cite{Fairlie:2008dg,Cachazo:2013gna}.

In this work we show that there are special subregions of ${\cal K}_n$ of dimension $n(n-3)/2$ on which all the solutions to the scattering equations are real. These regions also have the property that their co-dimension one boundaries are only collinear singularities. In order to define one such region we start by selecting three labels, say $\{A,B,C\}$, and declaring that the region is spanned by requiring the following $n(n-3)/2$ variables: $s_{ab}$, $s_{aA}$, $s_{aB}$ with $a,b$ not in $\{A,B,C\}$ to be non-negative. From this point on, we use the notation in which $a,b \in \{1,2,\ldots,n-3\}$ and $i,j \in \{1,2,\ldots,n\}$ in order to distinguish between the two sets. In section \ref{sec:geometry} we show that in such a region no multi-particle singularities, including those that involve particles from the set $\{A,B,C\}$, are possible. We denote this ``positive region'' by ${\cal K}_n^+$.

The moduli space of $n$-punctured spheres also has special subregions. Consider the subspace where all $n$ points lie on a circle which can then be mapped to the real line. Using ${\rm SL}(2,\mathbb{R})$ one can fix three punctures, say $\{\sigma_A,\sigma_B,\sigma_C\}$. Moreover, restricting the remaining $n-3$ points to be located in between $\sigma_A$ and $\sigma_B$, one has the configuration space of $n-3$ points on an interval $\mathtt{I}=[\sigma_A,\sigma_B]$. Let us denote this space as ${\rm Conf}_{n-3}(\mathtt{I})$.

\begin{figure}[!t]
	\centering
	\includegraphics[width=0.6\textwidth]{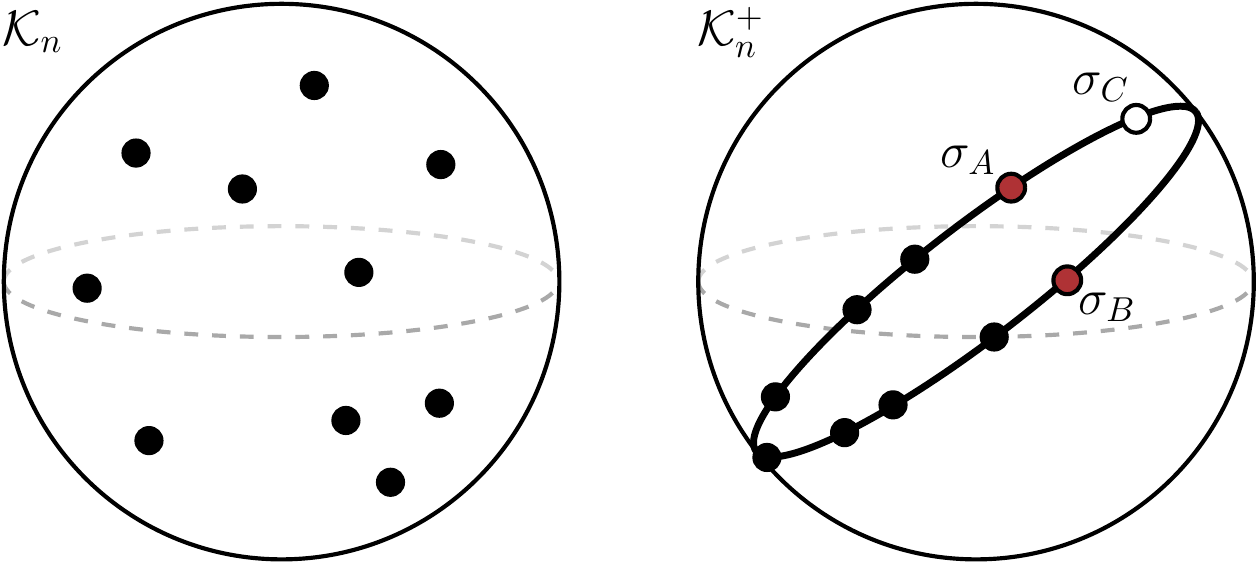}
	\caption{\label{fig:sphere}From $\mathbb{CP}^1$ to $\mathbb{RP}^1$. Left: Generic kinematics, $\mathcal{K}_{n}$ leads to configurations of punctures living on a Riemann sphere. Right: Positive kinematics, $\mathcal{K}^+_{n}$ makes the punctures arrange themselves on a real projective line. The black punctures $\sigma_1,\sigma_2,\ldots,\sigma_{n-3}$ can be thought of as a system of mutually repelling particles on the interval $\mathtt{I}=[\sigma_A,\sigma_B]$. Punctures $\sigma_A, \sigma_B, \sigma_C$ can be fixed using ${\rm SL}(2,\mathbb{R})$ invariance. The first two then provide repelling boundary conditions for the black particles.}
\end{figure}

In this work we find an intimate relation between ${\cal K}_n^+$ and ${\rm Conf}_{n-3}(\mathtt{I})$ via the scattering equations. In section \ref{sec:SE} we show that the scattering equations for a point in ${\cal K}_n^+$ become the equilibrium conditions for a system of $n-3$ mutually repelling particles on a real interval, where the end points also repel the $n-3$ particles, see figure \ref{fig:sphere}. This picture makes it obvious that for any of the $(n-3)!$ possible orderings of points on the interval, there exists at least one stable equilibrium point. It is well-known that the scattering equations possess $(n-3)!$ solutions \cite{Cachazo:2013gna} (we review this fact in appendix \ref{sec:Counting solutions}) and therefore for any point on ${\cal K}_n^+$ one can conclude that all solutions lie on ${\rm Conf}_{n-3}(\mathtt{I})$. This phenomenon was first observed by Kalousios \cite{Kalousios:2013eca} for a special two-dimensional subspace of ${\cal K}_n$, which gives rise to a beautiful connection to the Jacobi polynomials $P^{(\alpha,\beta)}_{n-3}(z)$. In fact, the kinematic space found by Kalousios is contained inside ${\cal K}_n^+$. We briefly review the connection to the Jacobi polynomials in appendix \ref{sec:Jacobi}.

In section \ref{sec:four dimensions} we specialize to four dimensions and start by showing that there are subspaces of ${\cal K}_n^+$ that satisfy the extra constraints needed to represent $n$ massless particles scattering in four dimensions, although only in split signature. In this dimension, the $(n-3)!$ solutions to the scattering equations are known \cite{Cachazo:2013iaa} to separate into $n-3$ sectors classified by a number $k\in \{2,3,\ldots ,n-2\}$. When amplitudes of particles with non-zero helicity, say $h=\pm 1$, are computed, one finds that $n$-particle amplitudes with $k$ negative helicity particles can be computed using solutions exclusively in the $k$-sector \cite{Witten:2003nn,Roiban:2004yf,Mason:2013sva,Geyer:2014fka,Du:2016fwe}. Moreover, it is also known \cite{Spradlin:2009qr}\cite{Cachazo:2013iaa} that the number of solutions in the $k$ sector is given by the Eulerian number $\eulerian{n-3}{k-2}$. A well-known combinatorial interpretation of the Eulerian number $\eulerian{n-3}{k-2}$ is that it counts the number of permutations of $(n-3)!$ elements with exactly $k-2$ descents. We find that this picture is exactly realized in our map from ${\cal K}_n^+$ to ${\rm Conf}_{n-3}(\mathtt{I})$. More precisely, we prove that if the unique solution corresponding to $k=2$, where $\eulerian{n-3}{0}=1$, is used to define the identity permutation, then solutions corresponding to permuting punctures to get $k-2$ descents are the ones in the $k$ sector.

In section \ref{sec:applications} we end with some applications and future directions. These include computation of amplitudes, efficient numerical algorithms, a vector space structure in ${\cal K}_n^+$ that leads to a novel structure on ${\rm Conf}_{n-3}(\mathtt{I})$ and other interesting regions with physical kinematics and applications.

\section{\label{sec:geometry}The Geometry of ${\cal K}_n^+$}

The space ${\cal K}_n^+$ is a special $n(n-3)/2$-dimensional subregion of the space of kinematic invariants $s_{ij}$ for the scattering of $n$ massless particles in which, as shown in the next section, the scattering equations have all real solutions. In this section we give an explicit construction and study some of its properties. Let us single out three particle labels, say $n-2$, $n-1$ and $n$ and call them $A$, $B$ and $C$ respectively. The property to be proven is that the space defined by $s_{ab}>0$, $s_{aA}>0$ and $s_{aB}>0$ with $a\in \{1,2,\ldots ,n-3\}$ has no multi-particle singularities.

Let us start by writing all other kinematic invariants in terms of the basis $\{ s_{ab}, s_{aA},s_{aB} \}$. Note that these are
$(n-3)(n-4)/2 + 2(n-3) = n(n-3)/2$ independent variables. All the remaining $n$ Mandelstam invariants are given by
\ba\label{others}
s_{aC} &=& - s_{aA}-s_{aB}-\sum_{b=1}^{n-3}s_{ab}, \quad\quad\quad s_{AB} = -\sum_{a=1}^{n-3}(s_{aA}+s_{aB})-\!\!\!\sum_{1\leq a< b \leq n-3}s_{ab},\tr
s_{AC} &=& \sum_{a=1}^{n-3}s_{aB} +\!\!\!\sum_{1\leq a< b \leq n-3}s_{ab}, \quad\quad\; s_{BC} = \sum_{a=1}^{n-3}s_{aA} +\!\!\!\sum_{1\leq a< b \leq n-3}s_{ab}.
\ea
Clearly, on ${\cal K}_n^+$ one has $s_{aC}<0$, $s_{AB}<0$ while $s_{AC}>0$ and $s_{BC}>0$.

All this information can be neatly summarized in the structure of the matrix of Maldelstam invariants with entries $s_{ij}$,
\be
\begin{array}{cc}
\begin{array}{cccccccc}
{\text{\tiny 1 \phantom{i}}} & {\text{\tiny 2 \phantom{i}}} & {\text{\tiny 3 \phantom{i}}} & \text{\tiny\phantom{i}$\cdots$\phantom{i}} & {\text{\tiny n-3\phantom{ii} }} & \text{\phantom{}\tiny A} & \text{\tiny \phantom{ii}B} & \text{\tiny \phantom{ii}C}\\
\end{array} & \\

\left[\begin{array}{@{}ccccc|c|c|c@{}}
\;0\; & \tikzmark{A}{south east} &  & \phantom{iii} & \phantom{iii} &  & \tikzmark{D}{south} &  \\
 & 0  &   & \textbf{+} &   & \textbf{+}  & \textbf{+}  & -  \\
 & \phantom{iii}  & {\scriptsize \ddots} &   &   &   &   &  \\
 & +  &   & 0  &   &   &   &   \\
 &   &   &   &  0 & \tikzmark{B}{north}  & \tikzmark{C}{north west}  &   \\
\hline
 & +  &   &   &   & 0 & - & + \\
\hline
 & +  &  &   &   & - & 0 & + \\
\hline
 & -  &   &   &   & + & + & 0 \\
\end{array}\right]

\begin{tikzpicture}[overlay,remember picture]
\draw[Maroon, thick, dashed] (A.north west) -- (B.center) -- ([xshift=2]C.center) -- ([xshift=2]D.north east) -- cycle;
\end{tikzpicture}

 & \begin{array}{l}
\text{\tiny 1}\\
\text{\tiny 2}\\
\text{\tiny 3}\\
\text{\tiny $\vdots$}\\
\text{\tiny n-3}\\
\text{\tiny A}\\
\text{\tiny B}\\
\text{\tiny C}\\
\end{array}\\
\end{array},
\en
where the elements in the basis are encircled by dashed lines.

A multi-particle factorization is defined by a subset ${\cal I}$ of $\{1,2,\ldots ,n-3,A,B,C\}$ with at least three elements (and less than $n-2$ elements by momentum conservation). We would like to study the quantity
\be\label{master}
\sum_{i,j\,\in \,{\cal I}}s_{ij}
\ee
and show that it can never be zero in the interior of ${\cal K}_n^+$. There are only two classes to consider, as all the other cases follow from momentum conservation. The first class is when ${\cal I}\subset \{1,2,\ldots ,n-3,A\}$ or ${\cal I}\subset \{1,2,\ldots ,n-3,B\}$, which are trivial since all the Mandelstam invariants involved are a part of the positively-defined basis. The second class is when $A,B\in {\cal I}$, but not $C$. Here we write \eqref{master} as
\be
\sum_{a,b\in {\cal I}\setminus\{A,B\}}\!\!\!\!\!s_{ab} + 2\!\!\!\!\!\!\sum_{a\in {\cal I}\setminus\{A,B\}}\!\!\!(s_{aA}+s_{aB}) + 2 s_{AB}.
\ee
Using \eqref{others} to write $s_{AB}$ in terms of the basis one finds
\be
-\left(\sum_{a,b=1}^{n-3}s_{ab}\; -\!\!\!\!\sum_{a,b\in {\cal I}\setminus\{A,B\}}\!\!\!\!\!s_{ab}\right) - 2\left(\sum_{a=1}^{n-3}\;(s_{aA}+s_{aB})\; - \!\!\!\sum_{a\in {\cal I}\setminus\{A,B\}} \!\!\!\!\!(s_{aA}+s_{aB})\right).
\ee
Since in both brackets every element in the second sum appears in the first, one finds that the answer is minus the sum of the elements of the basis, and thus can never vanish on ${\cal K}_n^+$.

Given that all co-dimension one boundaries of ${\cal K}_n^+$ are collinear singularities, it is natural to consider them in detail. We postpone this to appendix \ref{sec:collinear}, where the discussion also uses results from the next sections.

Before ending this section it is important to mention that there are other subregions of ${\cal K}_n$ with no multiparticle singularities that are not in ${\cal K}_n^+$. In fact, some of them correspond to the physical region for the scattering of $2\to n-2$ particles in Minkowski spacetime\footnote{We thank Lance Dixon for pointing this out to us.} as shown in \cite{Dixon:2016epj}. One can show that the scattering equations generically possess complex solutions in this region. However, in section \ref{sec:applications} we discuss a special sub-region where solutions are real and discuss some physical applications.

\section{\label{sec:SE}Scattering Equations on ${\cal K}_n^+$}

In this section we show that for any kinematic invariants belonging to ${\cal K}_n^+$ the scattering equations have all $(n-3)!$ real solutions (the counting of solutions in the general case is reviewed in appendix \ref{sec:Counting solutions}). It is particularly convenient to use the ${\rm SL}(2,\mathbb{C})$ redundancy to fix the punctures: $\sigma_A = 0$, $\sigma_B=1$ and $\sigma_C=\infty$. Of course, all the discussion below can be done in any gauge choice.

The scattering equations can be thought of as determining the extrema of the following potential:
\be
V(\sigma ) =\; -\!\!\!\sum_{1\leq a<b\leq n-3}s_{ab}\log |\sigma_a-\sigma_b|\; -\;\sum_{a=1}^{n-3}s_{aA} \log |\sigma_a| \;-\; \sum_{a=1}^{n-3}s_{aB} \log |1-\sigma_a|.
\ee
Note that all dependence on $\sigma_C$ has dropped out. Also, the term containing $s_{AB}$ is not present since $|\sigma_A-\sigma_B|=1$. Here the kinematic invariants are the coupling constants among particles. We interpret $V(\sigma)$ as the potential for $n-3$ particles moving in the real interval $\mathtt{I}=[0,1]$. In other words, we will be looking for solutions to the scattering equations that have this property. If we can show that all $(n-3)!$ solutions to the scattering equations are of this form then the proof of the reality of the solutions will be completed.

In order to prove that our picture is correct, note that all the coupling constants $s_{ab}$, $s_{aA}$, $s_{aB}$ appearing in the potential $V(\sigma)$ are precisely the elements in the basis of ${\cal K}_n^+$ and hence are positive. Since particles are restricted to be in the interval $\mathtt{I}$, all terms of the form $\log |\sigma_a-\sigma_b|$ are negative, including the interactions with the walls at $0$ and $1$.

The physical picture is then that of $n-3$ mutually repelling particles on an interval with walls at $0$ and $1$ that also repel the $n-3$ particles.

The last observation is that the potential blows up when two particles coalesce or a particle approaches any of the two walls. This means that once the order of the $n-3$ particles is chosen on the interval, they cannot change places. Therefore, it is clear that for each of the possible orderings of particles there exists {\it at least} one equilibrium point of $V(\sigma)$. Moreover, each such point is stable. The number of possible arrangements is clearly the number of permutations of $n-3$ points, i.e., $(n-3)!$. Since the scattering equations are known to have exactly $(n-3)!$ solutions \cite{Cachazo:2013gna}, this implies that for each ordering there is exactly one equilibrium point and hence it concludes the proof.

\section{\label{sec:four dimensions}Specializing to Four Dimensions}

In this section we find all subspaces in the interior of ${\cal K}_n^+$ that can be obtained from the scattering of $n$ massless particles in four dimensions. Next, we study how the separation of solutions to the scattering equations into sectors has an elegant realization for points in ${\cal K}_n^+$.

When $n>d+1$, the number of independent kinematic invariants is not $n(n-3)/2$ anymore, but it is reduced. In four dimensions the number of independent variables is $3n-10$. This reduction happens because all $5\times 5$ minors of the matrix of Maldelstam invariants must vanish. This is a set of highly non-linear polynomial equations. Luckily, there is a simpler way of proceeding; we use spinor-helicity variables to explicitly construct kinematics that lives in ${\cal K}_n^+$.

Each particle is described by a pair of spinors $\lambda_a,\tilde\lambda_a$. Here we take both spinors to be real and independent and therefore the spacetime has split signature, $\mathbb{R}^{2,2}$. We comment on other signatures below. The Lorentz group is ${\rm SL}(2,\mathbb{R})\times {\rm SL}(2,\mathbb{R})$ with each factor acting on only one of the spinors. The set of all $n$ spinors can be nicely arranged into two $2\times n$ matrices. Using Lorentz transformations it is possible to bring the spinors to the form
\bes\label{kine}
\Lambda &=& \Bigg(\begin{array}{@{}ccccccccc@{}}
	\;t_1 x_1 & \;t_2 x_2 & \;t_3 x_3 & \;\cdots & \;t_{n-3} x_{n-3} & \; 0 & \;\; t_B & \;\; t_C\;\\
	t_1     & t_2     & t_3     & \,\cdots & t_{n-3} & \;\,t_A     & \;\;t_B &\; 0\\
	\end{array} \Bigg),\tr
\tilde{\Lambda} &=& \Bigg(\begin{array}{@{}ccccccccc@{}}	\tilde{t}_1     & \tilde{t}_2     & \tilde{t}_3     & \,\cdots & \tilde{t}_{n-3} & \;\,\tilde{t}_A     & \;\,\tilde{t}_B & \;0\\
	\;\tilde{t}_1 \tilde{x}_1 & \;\tilde{t}_2 \tilde{x}_2 & \;\tilde{t}_3 \tilde{x}_3 & \;\cdots & \;\tilde{t}_{n-3} \tilde{x}_{n-3} & \; 0 & \;\; \tilde{t}_B & \;\; \tilde{t}_C\;\\
\end{array} \Bigg).
\ens
Given that Mandelstam invariants do not transform under the little group, we are free to set $t_A=t_B=t_C=1$ as well as $t_a=1$ for all $a\in \{1,2,\ldots ,n-3\}$.

One could proceed directly to study the constraints imposed by requiring the Mandelstam invariants to belong to ${\cal K}_n^+$. However, using the results of the previous section it is possible to streamline the discussion. It is known \cite{Fairlie:1972zz} that the scattering equations possess two very simple solutions\footnote{Here we use the standard notation, where $\langle ab\rangle$ is the determinant of the $2\times 2$ matrix constructed using columns $a$ and $b$ of $\Lambda$. The definition of $[ab]$ is analogous, but using $\tilde{\Lambda}$.},
\be
\label{eq:mhv1}
\sigma^{(1)}_a = \frac{\la aA \ra \la BC \ra}{\la aC \ra \la BA \ra}, \qquad \sigma^{(2)}_a = \frac{ [ aA ] [ BC ]}{[ aC ] [ BA ]}.
\en
Using the parametrization \eqref{kine} one finds that $\sigma^{(1)}_a= x_a$ while $\sigma^{(2)}_a =\tilde x_a$. This immediately means that $0<x_a<1$ and $0<\tilde{x}_a<1$. Moreover, a simple computation using \eqref{eq:mhv1} reveals that
\be\label{ord}
\left( x_a -x_b \right)\left(\tilde{x}_a -\tilde{x}_b \right) = \frac{s_{ab}\, s_{BC}\, s_{AC}}{s_{aC}\, s_{bC}\, s_{AB}}< 0.
\en
The last inequality follows from the restriction to be in ${\cal K}_n^+$. Therefore, whatever order the $x$'s are on the interval $[0,1]$, the $\tilde{x}$'s must be in the reverse. Without loss of generality one can relabel the points so that
\be\label{order}
0<x_1<x_2<\ldots < x_{n-4} < x_{n-3}<1, \qquad 0 < \tilde{x}_{n-3}< \tilde{x}_{n-4} < \ldots < \tilde{x}_2 < \tilde{x}_1 < 1.
\ee

Consider now the constraint that $s_{ab}=\langle ab\rangle[ab]=\tilde{t}_a\tilde{t}_b(x_a-x_b)(\tilde{x}_b-\tilde{x}_a) >0$. Using \eqref{ord} this leads to the requirement that $\tilde{t}_a\tilde{t}_b >0$ for all $a,b$ (note the order of the labels in \eqref{ord}). This is achieved by asking all $\tilde{t}_a$ to have the same sign.

Before completing the analysis of the conditions, let us impose momentum conservation on the data \eqref{kine}. It is simple to show that momentum conservation is equivalent to the following four constraints
\be
\tilde{t}_A = -\sum_{a=1}^{n-3}\tilde{t}_a(1-\tilde{x}_a),\quad \tilde{t}_B = -\sum_{a=1}^{n-3}\tilde{t}_a x_a,\quad \tilde{t}_C = \sum_{a=1}^{n-3}\tilde{t}_ax_a(1-\tilde{x}_a),\quad \sum_{a=1}^{n-3}\tilde{t}_a(x_a-\tilde{x}_a) = 0.\;
\ee
Note that the equations for $\tilde{t}_A$ and $\tilde{t}_B$, in combination with $\tilde{t}_a\tilde{t}_b >0$ for all $a,b$, imply the conditions $\tilde{t}_a\tilde{t}_B < 0$ and $\tilde{t}_a\tilde{t}_A < 0$ for all $a$.

Finally, the remaining constraints $s_{aB}= \langle aB\rangle[aB]=\tilde{t}_a\tilde{t}_B(x_a-1)(1-\tilde{x}_a) >0$ and $s_{aA}= \langle aA\rangle[aA]=\tilde{t}_a\tilde{t}_A(x_a)(-\tilde{x}_a) >0$ can be readily checked to be automatically satisfied.

One may wonder whether there are points in ${\cal K}_n^+$ with four-dimensional kinematics and other signatures. In addition to split signature one has Euclidean and Lorentzian. The former does not allow massless particles while the latter requires $(\lambda_a)^* = \pm \tilde\lambda_a$ and it is easy to show that the requirement is not compatible with positivity of the basis. However, in section \ref{sec:applications} we discuss regions outside of ${\cal K}_n^+$ which correspond to special physical scattering of $2\to n-2$ particles in Minkowski space which also lead to real solutions.

It is also important to mention that three dimensions is special in that there are no points in ${\cal K}_n^+$. One way to see this is by dimensional reduction of the four dimensional points. The momenta for particles $k_A^{\mu}$, $k_B^\mu$ and $k_C^\mu$ define a three dimensional space of the form $(k^0,k^1,k^2,0)$. Imposing that the $k^3$ components of all other particles vanish requires $\lambda_{1,a}\tilde\lambda_{1,a} = \lambda_{2,a}\tilde\lambda_{2,a}$, or using \eqref{kine} that $x_a = \tilde x_{a}$ which is clearly impossible for $n>4$ due to the ordering constraint \eqref{order}.

Lastly, let us briefly remark on a connection to the positive Grassmannian space \cite{ArkaniHamed:2012nw}. For this purpose, we need to rearrange the columns corresponding to the particle $A$ in \eqref{kine} to lie before particle $1$, i.e., we choose the ordering $(A,1,2,\ldots,n-3,B,C)$. One can then show that the momentum twistor $Z$ evaluated on the positive kinematics $\mathcal{K}_n^+$ always belongs to the positive Grassmannian, $Z \in G^+(4,n)$. In this regard, the space $\mathcal{K}_n^+$ is quite special, since general positive Grassmannian data produces complex solutions. We leave a more in-depth exploration of this connection for future research\footnote{There is a numerical evidence for general momentum twistors in the positive Grassmannian $G^+(4,n)$ to produce real solutions in the $k=3$ sector \cite{BourjailyPrivate} (in the notation of section \ref{sec:Separation into Sectors}).}.

\subsection{\label{sec:Separation into Sectors}Separation into Sectors}

Before continuing the discussion of points in ${\cal K}_n^+$, let us review why the solutions to the scattering equations are expected to separate into sectors for any kinematics in four dimensions. The scattering equations can be obtained by requiring the Lorentz vector
\be\label{eq:map}
P^\mu(\sigma) = \sum_{i=1}^n\frac{k^\mu_i}{\sigma-\sigma_i}
\ee
to be null for any value of $\sigma$ \cite{Cachazo:2013gna}. Therefore, on the support of the scattering equations it must be that
\be\label{eq:rational map}
\sigma_{\mu,\alpha\dot\alpha}P^\mu(\sigma ) = \frac{\lambda_\alpha(\sigma )\tilde\lambda_{\dot\alpha}(\sigma )}{\prod_{i=1}^n(\sigma-\sigma^\star_i)}
\ee
where $\sigma^\star_i$ denotes a specific solution, $\lambda_\alpha(\sigma )\tilde\lambda_{\dot\alpha}(\sigma )$ is a polynomial of degree $n-2$ and $\sigma_{\mu,\alpha\dot\alpha}$ is the $4$-vector of Pauli matrices. This structure implies that solutions can be classified by the degree of the polynomial $\lambda_\alpha(\sigma )$. A common notation is to denote ${\rm deg}\,\lambda_\alpha(\sigma )= k-1$ with $k\in \{2,3,\ldots, n-2\}$. Note that this choice fixes ${\rm deg}\,\tilde\lambda_{\dot\alpha}(\sigma )= n-k-1$.

It is also known \cite{Spradlin:2009qr,Cachazo:2013iaa} that the $(n-3)!$ solutions to the scattering equations separate so that $\eulerian{n-3}{k-2}$ of them belong to the $k$-sector. Here $\eulerian{n-3}{k-2}$ denotes an Eulerian number\footnote{The definition of Eulerian numbers is given recursively $\eulerian{p}{q} = (p-q)\eulerian{p-1}{q-1} + (q+1)\eulerian{p-1}{q}$ with $\eulerian{1}{0}=1$ and $\eulerian{p}{q}=0$ for $p\leq q$.}. The proof that Eulerian numbers appear in this problem proceeds by induction using the soft limit of a given particle. In this process, one can show that all solutions, ${\cal N}_{n,k}$, on sector $k$ for $n$ particles become either solutions in the same sector $k$ or in sector $k-1$ for $n-1$ particles. More explicitly, the number ${\cal N}_{n,k}$ splits exactly as $(k-1){\cal N}_{n-1,k}$ and $(n-k-1){\cal N}_{n-1,k-1}$. It is easy to show that these are the recursion relations for the Eulerian numbers with the correct boundary conditions, since ${\cal N}_{4,2}=\eulerian{1}{0}=1$.

Given the identification, the Eulerian numbers must satisfy $\sum_{k=2}^{n-2}\eulerian{n-3}{k-2} = (n-3)!$. In fact, this equation has a combinatorial meaning; permutations of $(n-3)$ elements can be classified according to the number of descents they possess. The Eulerian numbers $\eulerian{n-3}{k-2}$ count permutations with exactly $k-2$ descents. Next, we show how this is realized when the kinematic data is taken to lie in $\mathcal{K}^+_n$.

\subsection{Physical Realization of Eulerian Numbers}

Let us assume that a point in $\mathcal{K}^+_n$ corresponding to four dimensional kinematics has been chosen. The claim to be proven is that the separations into sectors of the $(n-3)!$ solutions to the scattering equations exactly coincides with the classification of permutations of $n-3$ particles on the interval $[0,1]$ by the number of descents.

In order to make the claim precise, one has to make a choice of the identity permutation. There are two canonical choices defined by the two special solutions presented in \eqref{eq:mhv1}. In fact, these are the solutions corresponding to the $k=2$ and $k=n-2$ sectors:
\be
\label{eq:mhv}
\sigma^{(k=2)}_a = \frac{\la aA \ra \la BC \ra}{\la aC \ra \la BA \ra}, \qquad \sigma^{(k=n-2)}_a = \frac{ [ aA ] [ BC ]}{[ aC ] [ BA ]}.
\en
We use the ordering given by the $k=2$ solution as the identity permutation and therefore it has, by definition, $k-2=0$ descents. This means that $k=n-2$ leads to the permutation with $k-2 = n-4$ descents, which is maximal, as seen from \eqref{order}.

In order to prove the claim for any sector $k$, we proceed again by induction assuming that the claim is known to be true for $n-1$ particles. Let us relabel the $n-3$ particles, so that the $k=2$ sector has the canonical ordering $0< \sigma^{(k=2)}_1 < \sigma^{(k=2)}_2 < \ldots < \sigma^{(k=2)}_{n-3} < 1$. In order to illustrate the task ahead consider the case $n=8$. Here  $(12345)$ is the $k=2$ solution, $(12354)$ has one descent $5\to 4$ and therefore it must belong to the $k=3$ sector, $(15243)$ has two descents $5\to 2$, $4\to 3$ so it must belong to the $k=4$ sector, etc.

Consider the soft limit of a given particle, say $m$, in the set $\{1,2,\ldots ,n-3 \}$. This corresponds to switching off all of the couplings, $s_{mi}\to 0$, between the soft particle and the rest of the particles, thus effectively leading to a system of scattering equations with one less particle. It is clear that a given solution of the $(n-3)$-particle problem corresponding to a given permutation with $k-2$ descents now has only two possibilities after the soft limit, it either becomes a permutation with $k-2$ descents or one with $k-3$ descents. In other words, if a label is removed, the number of descents can never decrease by more than one or increase. Of course, after any soft limit the identity permutation for $n-3$ particles becomes the identity permutation for $n-4$ particles, and hence the reference permutation does not change.

Note that even though in higher dimensions the permutations can still be classified into sectors based on the number of descents, they no longer correspond to the factorization of the scattering maps.

\section{\label{sec:applications}Future Directions}

The most natural future step is the classification of all the other regions of ${\cal K}_n$ giving rise to real solutions of the scattering equations. For instance, we can consider crossing one of the collinear singularities of ${\cal K}_n^+$. For this kinematics, there will exist multiple solutions with the same permutations. The only way for them to become complex is when all of the punctures align exactly in multiple solutions. Before this happens, the punctures must still lie on a real line, thus defining a finite kinematic neighborhood of ${\cal K}_n^+$ in which the solutions are real. It would be interesting to further characterize this space.

Let us now turn into other possible research directions.

\subsection{Amplitudes in the CHY Formulation}

One of the most immediate applications of the behavior of the scattering equations on $\mathcal{K}^+_n$ is that the Cachazo-He-Yuan (CHY) \cite{Cachazo:2013hca,Cachazo:2013iea,Dolan:2013isa,Cachazo:2014xea} formulas for scattering amplitudes literally become integrals over real variables supported by distributions,
\be\label{newF}
{\cal A}_n = \int_{[0,1]^{n-3}}\prod_{a=1}^{n-3}d\sigma_a\; \delta\left(E_a\right)\,{\cal I}_L(\sigma )\,{\cal I}_{R}(\sigma )
\ee
with $E_a = \partial V(\sigma )/\partial \sigma_a$ with the potential defined in section \ref{sec:SE}. Moreover, the hypercube $[0,1]^{n-3}$ can be separated into $(n-3)!$ regions defined by orderings of the coordinates such as $0< \sigma_1<\sigma_2<\ldots <\sigma_{n-3}<1$. Therefore, the evaluation of the amplitude splits into $(n-3)!$ integrals each containing a single critical point of $V(\sigma)$. Recall that the critical points are all stable. This means that the Jacobian is positive and therefore there is no need for any analytic continuation, as the delta functions can be taken to be true distributions.

These simple observations can have important consequences as one of the most obvious obstacles for connecting the CHY formulation to standard string theory formulas is the fact that in general CHY formulas had to be defined as multi-dimensional contour integrals which compute residues at complex points in the moduli space ${\cal M}_{0,n}$. In contrast, open string amplitudes are defined as integrals over real intervals. The new form (\ref{newF}) of the CHY integrals is very reminiscent of string theory formulas. It would be interesting to explore the connection to string theory formulas, when data is taken to be in $\mathcal{K}^+_n$. Of course, as it is also common practice in string theory, we can take the values of amplitudes obtained in $\mathcal{K}^+_n$ as the basis for analytic continuation to any other region in ${\cal K}_n$. Since the amplitudes are rational functions of the kinematic invariants, the analytic continuation is simply obtained by replacing the delta function distributions by their contour integral representation.

An interesting observation is that when the two half-integrands in the CHY formula are the same, i.e., ${\cal I}_L(\sigma )= {\cal I}_{R}(\sigma )$, the amplitude becomes a sum of $(n-3)!$ positive numbers for any kinematics in $\mathcal{K}^+_n$. Cases where this happens are amplitudes of gravitons and Galileons \cite{Cachazo:2014xea}. It is tempting to suggest that these amplitudes can be interpreted as computing volume of some sort.

\subsection{Numerical Evaluations}

The interpretation of the scattering equations as the equilibrium equations for a stable system of particles on an interval immediately suggests that one can complete the dynamics by adding a friction term, so that when the particles start in a given configuration (ordering) on the interval, their positions evolve in time $\sigma_a(t)$ according to
\be\label{eq:evolution}
\ddot{\sigma}_a=\sum_{\substack{b=1\\ b\neq a}}^{n-3}\frac{s_{ab}}{\sigma_a-\sigma_b}+\frac{s_{aA}}{\sigma_a}+\frac{s_{aB}}{\sigma_a-1}-\gamma_a \dot{\sigma}_a \quad{\rm with} \quad \gamma_a > 0,
\ee
until they reach the equilibrium point. The convergence rate to the equilibrium point depends on the particular choice of $\gamma_a$. It would be interesting to explore efficient algorithms for these evaluations. We have been able to find solutions for specific orderings for numbers of particles as large as $n=60$.

Another use of this kind of algorithms is that in four dimensions one can hand-pick the solutions corresponding to a given $k$ sector by simply selecting the initial conditions appropriately to permutations with $k-2$ descents after the $k=2$ configuration has been identified.

The form of (\ref{eq:evolution}) is also similar to the ones appearing in the study of pseudospectra of normal matrices \cite{Tao2011}, where the eigenvalues of a matrix repel each other under a deformation of the matrix. It would be fascinating to investigate possible connections to the scattering equations.

\subsection{Vector Space Structure}

The scattering equations provide a map from $\mathcal{K}^+_n$ to the space of $n-3$ points on the interval, ${\rm Conf}_{n-3}(\mathtt{I})$. One can arrange the $r \equiv n(n-3)/2$ elements in the basis as an element of $(\mathbb{R^+})^r$. It is clear that any two points in $(\mathbb{R^+})^r$ related by a global rescaling give rise to the same point on ${\rm Conf}_{n-3}(\mathtt{I})$ and therefore instead of $\mathcal{K}^+_n$ it is more appropriate to use the equivalence classes by thinking about it projectively. It is known that the collection of such points has the following vector structure space, see, e.g., \cite{TomLeinster}. Addition of two vectors $\textsf{v}_1 = \{s_{ab}^{(1)},s_{aA}^{(1)},s_{aB}^{(1)}\}$ and $\textsf{v}_2= \{s_{ab}^{(2)},s_{aA}^{(2)},s_{aB}^{(2)}\}$ is defined as term-wise multiplication, i.e.,
\be
\textsf{v}_1+\textsf{v}_2 = \{\,s_{ab}^{(1)}s_{ab}^{(2)},\, s_{aA}^{(1)}s_{aA}^{(2)},\, s_{aB}^{(1)}s_{aB}^{(2)}\,\}.
\ee
The zero element is clearly given by $\textsf{0} = \{ 1,1,\ldots ,1\}$. Finally, multiplication by a scalar $\rho\in \mathbb{R}$ of $\textsf{v} = \{s_{ab},s_{aA},s_{aB}\}$ is given by term-wise exponentiation, i.e., $\rho\,\textsf{v} = \{ s_{ab}^{\rho},s_{aA}^{\rho},s_{aB}^{\rho}\}$.

It would be interesting to explore how this structure induces a vector space on ${\rm Conf}_{n-3}(\mathtt{I})$ and we leave this for future research. Here we simply note that the zero element is a point in the family studied by Kalousios \cite{Kalousios:2013eca}, which we briefly summarize in appendix \ref{sec:Jacobi}. In his paper the following kinematics was considered: $s_{ab}=1$, $s_{aA} =(1+\beta)/2$ and $s_{aB}=(1+\alpha)/2$. Kalousios found that the solutions to the scattering equations are related to the roots of the Jacobi polynomials $P^{(\alpha,\beta)}_{n-3}(z)$ of degree $n-3$. The zero vector thus corresponds to the point $\alpha=\beta=1$, which coincidentally can be mapped to a certain integrable Calogero-Moser model \cite{Corrigan:2002th}.

\subsection{\label{sec:physical kinematics}Physical Kinematics and Applications}

\begin{figure}[!t]
	\centering
	\includegraphics[width=0.8\textwidth]{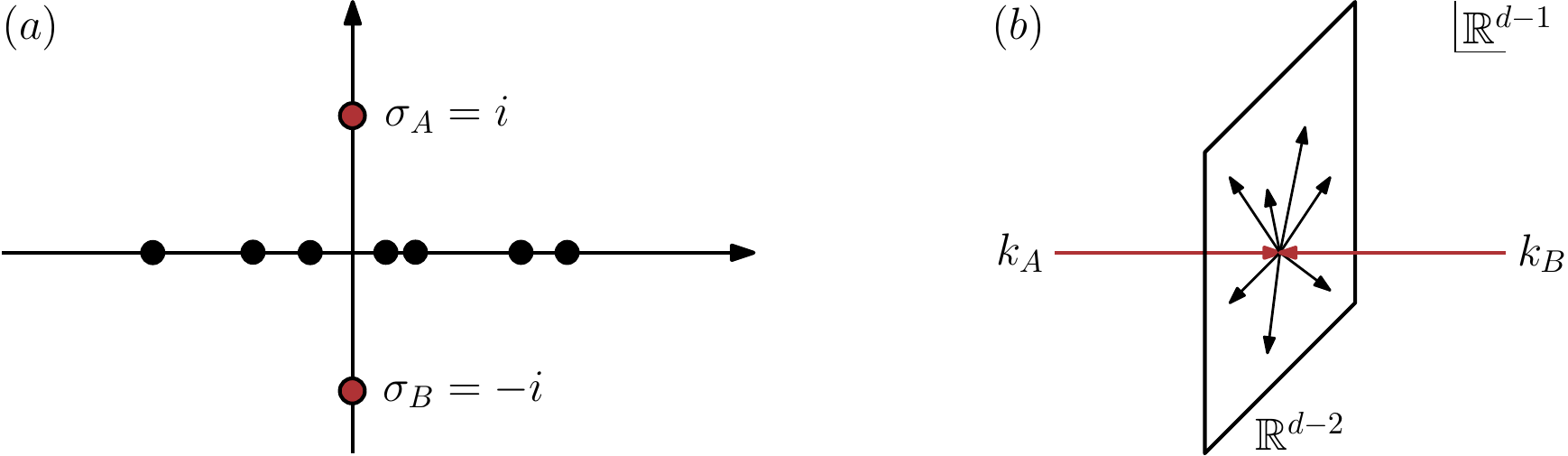}
	\caption{\label{fig:Minkowski-scattering}(a) Configuration of particles for physical kinematics from section \ref{sec:physical kinematics}. The vertical forces exerted on particles $1,2,\ldots,n-3$ due to $A$ and $B$ are equal and opposite, making them stay on the real axis. (b) Corresponding $2 \to n-2$ scattering process. Two massless particles of equal and opposite spatial momenta collide to produce $n-2$ particles confined to $d-2$ dimensions of space.}
\end{figure}

As mentioned at the end of section \ref{sec:geometry}, there is a physical region of the space of kinematics invariants which is bounded by only collinear limits. In other words, no multi-particle factorizations are present. This is the region for $2\to n-2$ scattering. While generic points in this region lead to complex solutions to the scattering equations, there is an interesting subspace where all solutions are real. Let the two incoming particles be $A$ and $B$. The special region is given by $s_{ab}>0$, $s_{aA}=s_{aB}<0$ and additional constraints we want to study. In order to make the behavior of the solutions clear, it is useful to use ${\rm SL}(2,\mathbb{C})$ transformations to set $\sigma_A = i$, $\sigma_B = -i$ and $\sigma_C = \infty$. Here $i^2=-1$. It is easy to see from figure \hyperref[fig:Minkowski-scattering]{2(a)}, that if $|s_{aA}| \gg |s_{ab}|$ then all particles $a\in \{1,2,\ldots , n-3\}$ will be trapped on the real line in a region around $\sigma=0$, due to the attraction from particles $A$ and $B$. Since particles $a,b$ repel each other, there clearly are $(n-3)!$ configurations corresponding to all ordering of the particles on the line.

Let us construct explicit momentum vectors that realize this kinematics. This will also hint at possible applications. For this part of the discussion we work in arbitrary number of dimensions, $\mathbb{R}^{d-1,1}$. One can always choose a frame where
\be
k_A = (-E,0,0,\ldots ,0,E), \quad k_B = (-E,0,0,\ldots ,0,-E).
\ee
Here we take the momentum vectors of $A$ and $B$ to be ingoing and therefore $E>0$. In order to impose that $s_{aA}=s_{aB}$ it must be that
\be
k_a = (k_a^0,k_a^1,\ldots , k_a^{d-2}, 0) \quad {\rm for} \quad a\in \{1,2,\ldots, n-3\}.
\ee
while $k_C =(k_C^0,k_C^1,\ldots , k_C^{d-2}, 0)$ is determined by momentum conservation. Here $s_{aA}=s_{aB} = - 2 E k_a^0$. The fact that $s_{ab}>0$ is a simple consequence of vectors being on-shell, i.e. $k^2_a=0$, future directed, i.e. $k_a^0>0$, and of the triangle inequality for the spatial part of the vectors.

Starting from the region in which $E \gg |k_a|$ and lowering $E$, we can now ask what is the critical value of $E$ at which the solutions first become complex. Note that for large $E$ the kinematic invariants
\be
s_{aC} = 4 E k_a^0 - \sum_{b=1}^{n-3} s_{ab}
\en
are always positive. In fact, the inequalities $s_{aC}>0$, $a \in \{1,2,\ldots,n-3\}$ define a natural region of the kinematic space that shares many similarities with the positive kinematics $\mathcal{K}_n^+$ studied in this work. This time, it is $\sigma_C = \infty$ that provides repelling boundary conditions for the remaining particles, while the interaction with $\sigma_A = i$ and $\sigma_B = -i$ keep them confined to the real axis. These conditions guarantee the existence of $(n-3)!$ stable equilibrium points on the real axis. In similarity to $\mathcal{K}_n^+$, the Jacobian is always positive.

However, we find that even if $s_{aC}<0$, the solutions to the scattering equations can still be real. It would be interesting to further characterize the critical point of $E$ at which the solutions first start becoming complex.

This special region has particles $A$ and $B$ approaching each other along the $(d-1)^{\rm th}$ direction and all other $n-2$ particles being produced purely in the transverse directions, see figure \hyperref[fig:Minkowski-scattering]{2(b)}. There are several immediate applications. The first is to consider $k_A+k_B =(-2E, 0,0,\ldots, 0)$ as the momentum vector of a massive particle with mass $M=2E$ in its center of mass frame in $\mathbb{R}^{d-2,1}$ rather than $\mathbb{R}^{d-1,1}$. This means that we can use the potential function
\be\label{eq:potential-5.4}
V(\sigma ) =\; -\!\!\!\!\!\sum_{1\leq a<b\leq n-3} \!\!\! s_{ab}\log |\sigma_a-\sigma_b|\; + \; M \sum_{a=1}^{n-3}k_a^0\log |1+\sigma^2_a|.
\ee
to study the decay of a massive particle into $n-2$ massless ones. Another application worth mentioning is to the computation of form factors where $P^\mu =k_A^\mu+k_B^\mu$ is defined as the off-shell momentum.

Let us take an example for $d=4$ in the special subregion $s_{aC}>0$ discussed above. It can be shown the most general parametrization of spinors is given by
\bes
\Lambda &=& \Bigg(\begin{array}{@{}cccccccc@{}}
	\;\,x_1 & \;\;\;\,x_2 & \;\;\dots & \;\;\;\,x_{n-3} & \;\;\;\;\;\;i & \;\;\;\;-i & \;\;\;\;\;\, i\;\;\;\\
	\;\,1   & \;\;\;1     & \;\;\dots & \;\;\;\,1       & \;\;\;\;\;\;1 &  \;\;\;\;\;1 &\;\;\;\;\;\, 0\;\;\;\\
\end{array} \Bigg),\tr
\tilde{\Lambda} &=& \Bigg(\begin{array}{@{}cccccccc@{}}	\;\tilde{t}_1x_1    &\tilde{t}_2 x_2   & \dots     & \tilde{t}_{n-3} x_{n-3}    & -i\tilde{t}_A&  \;i\tilde{t}_B & \;-i t_C\\
	\; \tilde{t}_1& \;\tilde{t}_2 & \;\dots & \;\tilde{t}_{n-3} & \;\tilde{t}_A &\;\tilde{t}_B & \;0\\
\end{array} \Bigg).
\ens
with
\bes
x_1<x_2<\dots<x_{n-3},&&\qquad \tilde{t}_a \tilde{t}_b>0,\qquad \tilde{t}_a \tilde{t}_C>0,\tr \tilde{t}_A=\tilde{t}_B=-\frac{1}{2}\sum_{a=1}^{n-3}\tilde{t}_a ,&&\qquad \tilde{t}_C=\sum_{a=1}^{n-3}\tilde{t}_a(1-x_a^2).
\ens
Two solutions of the scattering equations are given by
\be
\sigma^{(k=2)}_{a}=x_a, \qquad \sigma^{(k=n-2)}_{a}=-x_a.
\en
In fact, the potential \eqref{eq:potential-5.4} has a flip symmetry $\sigma_a \leftrightarrow -\sigma_a$ for all particles at the same time. Because of this fact, different branches of solutions are related to each other as $\sigma^{(k)}_a = - \sigma^{(n-k)}_a$. When this process is thought of as a decay of a massive particle or a form factor in three dimensions, the scattering equations inherit this structure.

It would be interesting to see how the case $d=5$ fits with the recently-found connected formulas for form factors in four dimensions \cite{He:2016jdg,Brandhuber:2016xue}. We leave this question for future research.

\section*{Acknowledgements}

We would like to thank Chrysostomos Kalousios for collaboration at an early stage of this project. We also thank N. Berkovits, J. Bourjaily, S. He, E. Yuan, and especially L. Dixon for useful discussions. This research was supported in part by Perimeter Institute for Theoretical Physics. Research at Perimeter Institute is supported by the Government of Canada through the Department of Innovation, Science and Economic Development Canada and by the Province of Ontario through the Ministry of Research, Innovation and Science.

\renewcommand{\thefigure}{\thesection.\arabic{figure}}
\renewcommand{\thetable}{\thesection.\arabic{table}}
\appendix

\section{\label{sec:Counting solutions}Counting the Number of Solutions}

\begin{figure}[!t]
	\centering
	\includegraphics[width=\textwidth]{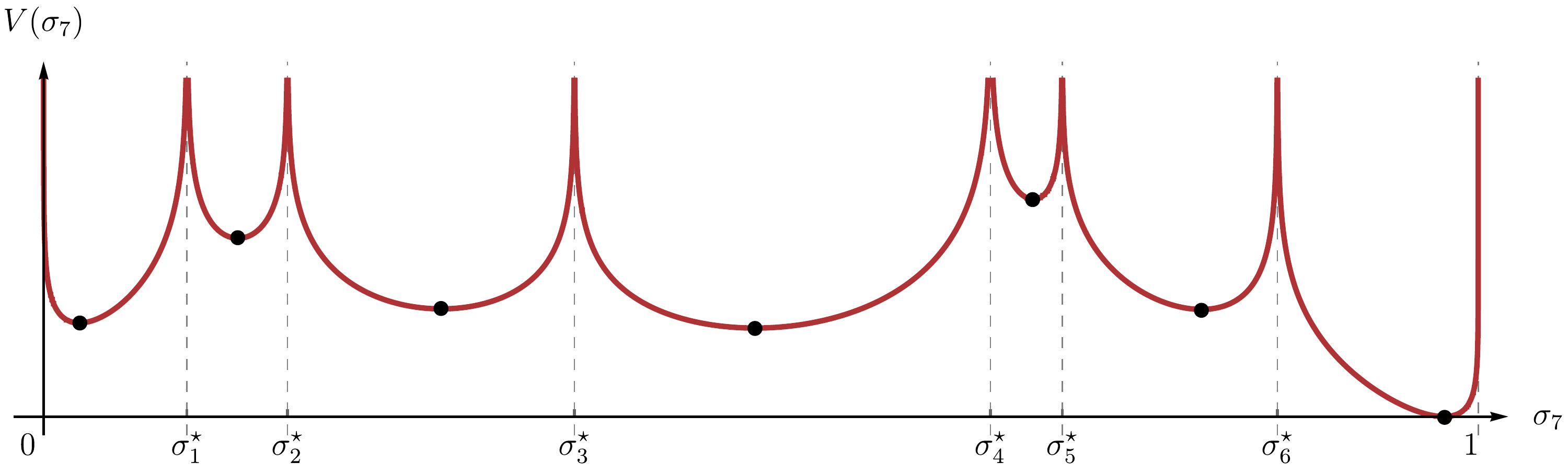}
	\caption{\label{fig:soft}Example of the potential, $V(\sigma_7)$, for the soft particle with label $7$. The minima of the potential mark the positions of the puncture $\sigma_7$ in different solutions.}
\end{figure}

In this appendix we review the derivation of the fact that the scattering equations have $(n-3)!$ solutions, originally found in \cite{Cachazo:2013gna}.

Firstly, let us consider the boundary case, $n=4$. Due to the ${\rm SL}(2,\mathbb{C})$ redundancy we have only one independent scattering equation, which is linear and hence has a single solution. We can now proceed with a proof by induction.

Let us take the $n^{\rm th}$ particle to be soft, i.e., $s_{in} = \tau \hat{s}_{in}$ with $\tau \to 0$ and $i \in \{1,2,\ldots, n-1\}$. The first $n-1$ scattering equations then become:
\be\label{eq:n-1 SE}
\sum_{\substack{j=1\\ j\neq i}}^{n-1}\frac{s_{ij}}{\sigma_i-\sigma_j} + {\cal O}(\tau) =  0 \quad {\rm for} \quad i\in\{ 1,2,\ldots ,n-1\}.
\ee
We know that in the soft limit these equations are identical to the ones for $n-1$ particles. By inductive assumptions there are exactly $(n-4)!$ of such solutions.

The last scattering equation becomes:
\be\label{eq:soft SE}
\tau \sum_{i=1}^{n-1}\frac{\hat{s}_{ni}}{\sigma_n-\sigma_i} =  0.
\ee
Hence for each solution of (\ref{eq:n-1 SE}) we obtain a polynomial constraint of degree $n-3$ for $\sigma_n$ (the leading order in $\sigma_n$ vanishes due to momentum conservation). Hence the total number of solutions is $(n-3)(n-4)! = (n-3)!$, which concludes the proof.

The above construction has an easy interpretation in terms of the particles living on an interval. Let us focus on a single solution of (\ref{eq:n-1 SE}). The soft particle experiences the potential created by the rest of the system through (\ref{eq:soft SE}), see figure \ref{fig:soft}. Since its contribution to (\ref{eq:n-1 SE}) is vanishing, it does not back-react on the environment. Hence the solutions of (\ref{eq:soft SE}) are given by the minima of the potential with all the other punctures $\sigma_i^\star$ fixed by (\ref{eq:n-1 SE}). Each minimum leads to a distinct permutation. It is interesting to observe how all the $(n-3)!$ permutations can be constructed by taking consecutive soft limits.

\section{\label{sec:Jacobi}Special Kinematics and Orthogonal Polynomials}

In this appendix we consider special cases of the positive scattering data ${\cal K}_n^+$, where a given number of particles are indistinguishable from each other from the perspective of the scattering equations. It was originally found by Kalousios that in this special kinematics, the solutions of the scattering equations are given by the roots of the Jacobi polynomials \cite{Kalousios:2013eca}. Here we briefly summarize this finding.

In the gauge fixing $(\sigma_A, \sigma_B, \sigma_C) = (-1,1,\infty)$ the scattering map (\ref{eq:map}) becomes,
\be
P^\mu (z) =  \sum_{a=1}^{n-3} \frac{k^\mu_a}{z-\sigma_a} + \frac{k_A^\mu}{z+1} +  \frac{k^\mu_B}{z-1}.
\en
In order to make the particles look the same to the scattering equations, we can set $s_{aA}=s_{bA}$ and $s_{aB}=s_{bB}$, as well as rescale all $s_{ab}=1$ for $a, b \in \{1,2,\ldots,n-3\}$. Such kinematics is always possible to construct in high enough dimension. Let us now assume that there exists a polynomial whose roots control the solutions of the scattering equations,
\be
\label{eq:polynomial}
Q(z) = \prod_{a=1}^{n-3} (z-\sigma_a).
\en
All $(n-3)!$ solutions are then related by permutations of these roots. Along the lines of \cite{Kalousios:2013eca}, we can consider the combination
\bes
2(1-z^2) Q(z) P^2(z) &=& (1-z^2) \sum_{a=1}^{n-3} \sum_{b = 1}^{n-3} \prod_{\substack{c=1\\ c \neq a,b}}^{n-3} (z-\sigma_c) - 4(z-1) \sum_{a=1}^{n-3} k_a \cdot k_A \prod_{\substack{b=1\\ b \neq a}}^{n-3}(z-\sigma_b) \tr
&& -\; 4(z+1) \sum_{a=1}^{n-3} k_a \cdot k_B \prod_{\substack{b=1\\ b \neq a}}^{n-3}(z-\sigma_b) - 4 k_A \cdot k_B \prod_{a=1}^{n-3}(z-\sigma_a)\tr
&=& (1-z^2)Q^{\prime\prime}(z) + 2\left(s_{aA}-s_{aB} -(s_{aA} + s_{aB})z\right) Q^{\prime}(z) - 2s_{AB} Q(z) =0,\nonumber
\ens
which vanishes by the requirement that the rational map is null everywhere. Since we already have fixed $(n-3)^2$ kinematic invariants, we are left with $2(n-3)$ degrees of freedom. The additional requirement that the $n-3$ particles need to be permutationally symmetric brings us to only two free parameters. The parametrization chosen in \cite{Kalousios:2013eca} reads:
\be
s_{aA} = \frac{1+\beta}{2},\quad s_{aB} = \frac{1+\alpha}{2}  \qquad\text{for}\qquad a \in \{1,2,\ldots,n-3\}.
\en
Solving for the remaining kinematics leads to the constraint equation for the polynomial $Q(z)=P^{(\alpha,\beta)}_{n-3}(z)$,
\be
(1-z^2) P_{n-3}^{(\alpha,\beta)\prime\prime}(z) + \left( \beta-\alpha - (2+\alpha+\beta)z \right) P^{(\alpha,\beta)\prime}_{n-3}(z) + n(n+\alpha+\beta+1) P^{(\alpha,\beta)}_{n-3} (z) = 0.\quad
\en
This is the differential equation defining the Jacobi polynomials. It follows from (\ref{eq:polynomial}) that the roots of these polynomials correspond to the solutions of the scattering equations. Kalousios showed that certain amplitudes evaluated on this kinematics can be obtained in a closed form \cite{Kalousios:2013eca}. He also showed that in the special case $\alpha=\beta=1$, the system is equivalent to a certain integrable Calogero-Moser model \cite{Corrigan:2002th}.

Let us comment on how the known results on the zeros of Jacobi polynomials fit nicely into the picture of particles interacting on an interval. Firstly, Jacobi polynomials are defined for $\alpha, \beta > -1$, which agrees with our positivity constraint on the kinematics. The boundary values correspond to the cases when one of the roots become $1$ and $-1$ respectively \cite{Olver:2010:NHMF}, which are the collinear limits. Secondly, the $n-3$ real roots are guaranteed to lie in the open interval $[-1,1]$ \cite{Olver:2010:NHMF}, precisely between the two fixed particles, $\sigma_A$ and $\sigma_B$. Lastly, different solutions of the scattering equations naturally arise from the permutations of the roots of the Jacobi polynomials.

It is possible to generalize the above derivation to the case of less than $n-3$ indistinguishable particles. One finds the answer can be expressed as the roots of Heine-Stieltjes polynomials \cite{Olver:2010:NHMF,ALRASHED1985327}.

\section{\label{sec:collinear}Collinear Factorization on ${\cal K}_n^+$}

It has been shown in section \ref{sec:geometry} that ${\cal K}_n^+$ does not have multi-particle singularities. In this appendix we study how the solutions behave as we approach a collinear factorization, say $s_{ab} \to 0$.

\begin{figure}[!t]
	\centering
	\includegraphics[width=.55\textwidth]{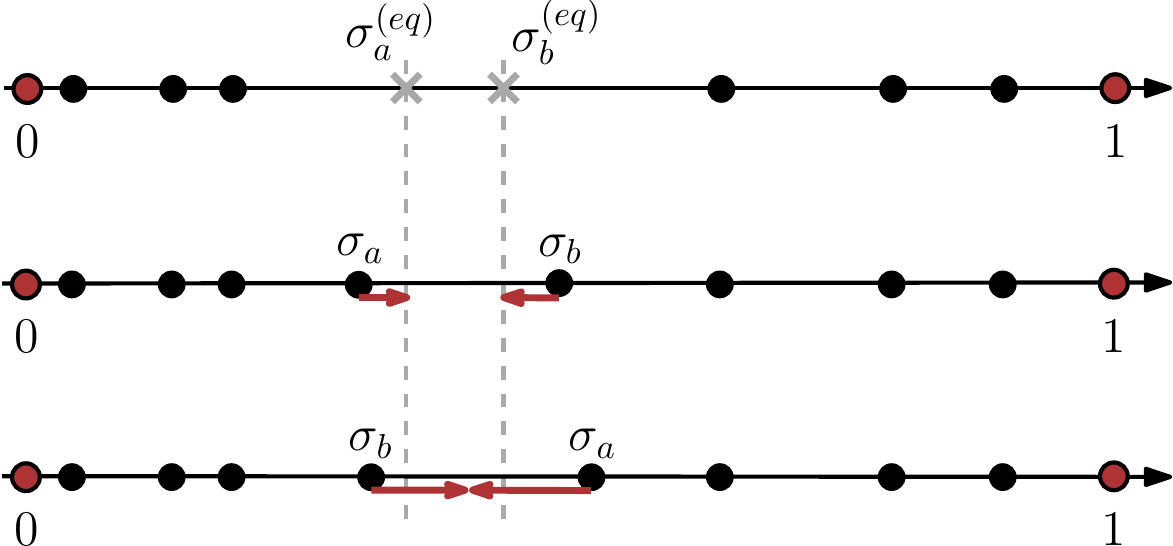}
	\caption{\label{fig:collinear}Determining which solution becomes degenerate. Top: Equilibrium positions that the particles try to reach. Middle: In the case $\sigma_a < \sigma_b$ the particles converge to distinct locations. Bottom: In the case $\sigma_b < \sigma_a$ the particles pinch at a common location creating a degenerate solution.}
\end{figure}

The only time when particles are allowed to pinch in a given permutation is when they are adjacent. There are $2(n-4)!$ such distinct configurations. Let us consider pairs of such solutions, in which all the remaining particles are in the same order, but one has $\sigma_a < \sigma_b$ and the other $\sigma_b < \sigma_a$. At the collinear point, i.e., $s_{ab}=0$, the interactions between particles $a$ and $b$ become vanishing, so nothing forbids the particles from meeting and passing through each other. In this case the system would have a natural equilibrium at points, either $\sigma_a^{\rm (eq)} < \sigma_b^{\rm (eq)}$ or $\sigma_b^{\rm (eq)} < \sigma_a^{\rm (eq)}$. It is enough to consider the first option, see figure \ref{fig:collinear}.

Let us start with finite $s_{ab}$ and approach the collinear limit. We find that the particles in the initial configuration $\sigma_a < \sigma_b$ can safely approach their respective equilibrium points, and thus stay finite in this limit. The opposite configuration however, $\sigma_b < \sigma_a$ will necessarily have particles pinching on their ways to $\sigma_a^{\rm (eq)} < \sigma_b^{\rm (eq)}$. For small but non-zero value of $s_{ab} >0$, the potential still blows up as they approach, so that they cannot cross and therefore coalesce. Since this behavior is the same for all the $(n-4)!$ adjacent pairs, we conclude that exactly $(n-4)!$ solutions will become degenerate on the collinear boundaries of ${\cal K}_n^+$. This is precisely the same answer one finds from the general analysis of the factorization properties of the scattering equations \cite{Cachazo:2013gna}.

\subsection{Four Dimensions}

In four dimensions, there are two ways of approaching the collinear limit, $\la ab \ra \to 0$ or $[ ab ] \to 0$. We will see how different permutations react differently to those limits, allowing us to distinguish between the two. As we have seen in section \ref{sec:four dimensions}, one can classify subsectors of ${\cal K}_n^+$ by a single ordering of $x$'s. Collinear factorization of the particles non-adjacent in this ordering are not reachable within ${\cal K}_n^+$, as it is evident from (\ref{order}).

We can classify the solutions depending on how they approach the collinear limit \cite{Roiban:2004yf}, say for particles $a$ and $b$:
\bes
&& \sigma_a - \sigma_b = \la ab \ra\, F(\{\lambda,\tilde{\lambda}\}),\tr
\text{or} && \sigma_a - \sigma_b = [ ab ]\; G(\{\lambda,\tilde{\lambda}\}),
\ens
Taking consecutive soft limits of the remaining particles from $\{1,2\ldots,n-3\}\setminus\{a,b\}$ can only change the functions $F$, $G$, but cannot discretely jump between the two cases. In the end, we are left with $5$ particles, which lie either in the $k=2$ or the $k=3$ sector, providing a natural identification with one of the two categories. We can thus conclude that for any permutation, whenever two adjacent particles have the same ordering as in the $k=2$ case, they become degenerate in the holomorphic collinear limit and stay finite in the anti-holomorphic limit. The opposite happens for the ordering not agreeing with $k=2$.

Let us illustrate this point with an example for $n=7$. Given the $k=2$ configuration $(1234)$, in the limit $[23] \to 0$ the permutations $(3214), (3241), (1324), (4321), (1432), (4132)$ become degenerate, in the limit $\la 34 \ra \to 0$ we have $(3412), (3421), (1342), (2341), (1234), (2134)$ contribute, and the limit $\la 13\ra \to 0$ is not reachable.

\bibliographystyle{JHEP}
\bibliography{references}

\end{document}